\documentclass[twocolumn,showpacs,preprintnumbers,amsmath,amssymb,superscriptaddress,nofootinbib]{revtex4}

\usepackage{dcolumn}
\usepackage{bm}

\usepackage{epsfig}
\usepackage{color}
\usepackage{graphicx}

\newcommand{\be}{\begin{equation}}
\newcommand{\ee}{\end{equation}}
\newcommand{\bea}{\begin{eqnarray}}
\newcommand{\eea}{\end{eqnarray}}
\newcommand{\ba}{\begin{array}}
\newcommand{\ea}{\end{array}}
\newcommand{\ket}[1]{\left\lvert #1\right\rangle}
\newcommand{\bra}[1]{\left\langle #1\right\rvert}

\long\def\symbolfootnote[#1]#2{\begingroup%
\def\thefootnote{\fnsymbol{footnote}}\footnote[#1]{#2}\endgroup} 

\begin{document}

\preprint{Caltech MAP-331}

\topmargin=0pt

\title{\Large
Pion Leptonic Decays and Supersymmetry}

\author{Michael J. Ramsey-Musolf}
\affiliation{California Institute of Technology, Pasadena, CA 91125}
\affiliation{University of Wisconsin, Madison, WI, 53706-1390}
\author{Shufang Su}
\affiliation{Department of Physics, University of Arizona, Tucson, AZ 85721}
\author{Sean Tulin} 
\affiliation{California Institute of Technology, Pasadena, CA 91125}

\date{June 4, 2007}

\begin{abstract}

We compute supersymmetric contributions to pion leptonic ($\pi_{l2}$) decays in the Minimal Supersymmetric Standard Model (MSSM).  When R-parity is conserved, the largest contributions to the ratio $R_{e/\mu} \equiv \Gamma[ \pi^+ \to e^+ \nu_e(\gamma)]/\Gamma[ \pi^+ \to \mu^+ \nu_\mu(\gamma)]$ arise from one-loop $(V-A)\otimes (V-A)$ corrections.  These contributions can be potentially as large as the sensitivities of upcoming experiments; if measured, they would imply significant bounds on the chargino and slepton sectors complementary to current collider limits.  We also analyze R-parity violating interactions, which may produce a detectable deviation in $R_{e/\mu}$ while remaining consistent with all other precision observables.  

\end{abstract}

\maketitle

\section{\label{sec:level1}Introduction}

Low-energy precision tests provide important probes of new physics that are complementary to collider experiments\cite{Erler:2004cx,Ramsey-Musolf:2006ur,Ramsey-Musolf:2006ik}.  In particular, effects of weak-scale supersymmetry (SUSY) --- one of the most popular extensions of the Standard Model (SM) --- can be searched for in a wide variety of low-energy tests: muon $(g-2)$~\cite{Martin:2001st}, $\beta$- and $\mu$-decay~\cite{Kurylov:2001zx,Profumo:2006yu}, parity-violating electron scattering~\cite{Kurylov:2003zh}, electric dipole moment searches~\cite{Pospelov:2005pr}, and SM-forbidden transitions like $\mu \to e \gamma$~\cite{lfv}, \textit{etc} (for a recent review, see Ref.~\cite{Ramsey-Musolf:2006vr}).  In this paper, we compute the SUSY contributions to pion leptonic  ($\pi_{l2}$) decays and analyze the conditions under which they can be large enough to produce observable effects in the next generation of experiments. 

In particular, we consider the ratio $R_{e/\mu}$, defined by 
\begin{equation}
R_{e/\mu} \equiv \frac{\Gamma(\pi^+ \to e^+ \nu_e + e^+ \nu_e \gamma) }
                      {\Gamma(\pi^+ \to \mu^+ \nu_\mu+\mu^+ \nu_\mu \gamma) } \;.
\end{equation}
The key advantage of $R_{e/\mu}$ is that a variety of QCD effects that bring large theoretical uncertainties--- such as the pion decay constant $F_\pi$ and lepton flavor independent QCD radiative corrections --- cancel from this ratio.  Indeed, $R_{e/\mu}$ is one of few electroweak observables that involve hadrons and yet are precisely calculable (see~\cite{Bryman:1993gm} for discussion and Refs.~\cite{Marciano:1993sh,Decker:1994ea} for explicit computations).  Moreover, measurements of this quantity provide unique probes of deviations from lepton universality of the charged current (CC) weak interaction in the SM that are induced by loop corrections and possible extensions of the SM. In the present case, we focus on contributions from SUSY that can lead to deviations from lepton universality. 

Currently, the two most precise theoretical calculations of $R_{e/\mu}$ in the SM are~\cite{Marciano:1993sh, Decker:1994ea}  
\be
\label{eq:remusm}
R^{SM}_{e/\mu} = \left\{ \begin{array}{c} (1.2352 \; \pm \; 0.0005) \times 10^{-4}\\
                         (1.2356 \; \pm \; 0.0001) \times 10^{-4} \end{array} \right.
\ee
where the theoretical uncertainty comes from pion structure effects.  By utilizing chiral perturbation theory, it may be possible to reduce this uncertainty even further~\cite{futurestudy}.
Experimentally, the most precise measurements of $R_{e/\mu}$ have been obtained at  TRIUMF~\cite{Britton:1992xv} and PSI~\cite{Czapek:1993kc}.  Taking the average of these results gives~\cite{Yao:2006px}
\begin{equation}
R^{EXPT}_{e/\mu} = (1.230 \; \pm \; 0.004 ) \times 10^{-4} \;,
\end{equation}
in agreement with the SM.  Future experiments at these facilities will make more precise measurements of $R_{e/\mu}$, aiming for precision at the level of $\: < 1\times 10^{-3}$ (TRIUMF~\cite{triumfproposal}) and $5\times 10^{-4}$ (PSI~\cite{psiproposal}).  These projected uncertainties are close to the conservative estimate of theoretical uncertainties given in Ref.~\cite{Marciano:1993sh}.

\begin{figure*}[!ht]
\begin{center}
\mbox{\hspace*{-0.7cm}\epsfig{file=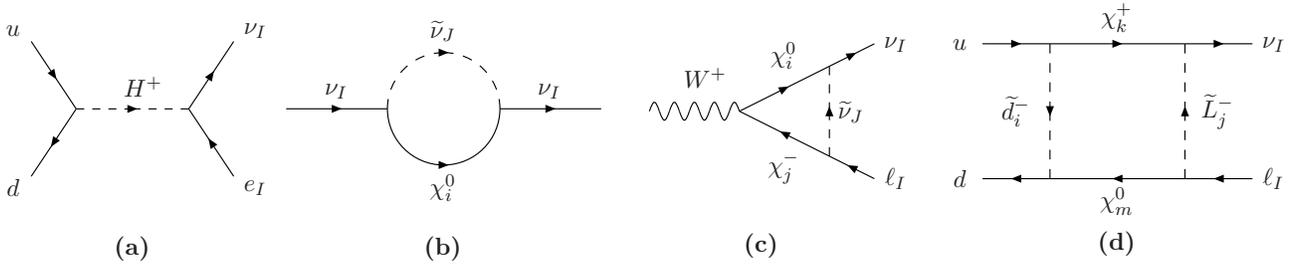,height=3.5cm}}
\end{center}
\caption{\it\small 
Representative contributions to $\Delta R^{\textrm{SUSY}}_{e/\mu}$: (a) tree-level charged Higgs boson exchange, (b) external leg diagrams, (c) vertex diagrams, (d) box diagrams.  Graph (a) contributes to the pseudoscalar amplitude, graphs (b,c) contribute to the axial vector amplitude, and graph (d) contributes to both amplitudes.
}
\label{fig:feynman}
\end{figure*}

Deviations $\Delta R_{e/\mu}$ from the SM predictions in Eq.~(\ref{eq:remusm}) would signal the presence of new, lepton flavor-dependent physics. In the Minimal Supersymmetric Standard Model (MSSM), a non-vanishing $\Delta R_{e/\mu}^{SUSY}$ may arise from either tree-level or one-loop corrections.  In section \ref{sec:loops}, we consider contributions to $\Delta R_{e/\mu}^{SUSY}$ arising from R-parity conserving interactions (Fig.~\ref{fig:feynman}). Although tree-level charged Higgs exchange can contribute to the rate $\Gamma[\pi^+\to\ell^+\nu(\gamma)]$, this correction is flavor-independent and cancels from $R_{e/\mu}$.
One-loop corrections induce  both scalar and vector semileptonic dimension six four-fermion operators. Such interactions contribute via pseudoscalar and axial vector pion decay matrix elements, respectively.  We show that the pseudoscalar contributions are negligible unless the ratio of the up- and down-type Higgs vacuum expectation values (vevs) is huge ($v_u/v_d \equiv \tan \beta \gtrsim 10^3$). For smaller $\tan\beta$ the most important effects arise from one-loop contributions to the axial vector amplitude, which we analyze in detail by performing a numerical scan over MSSM parameter space. We find that experimental observation of SUSY loop-induced deviations at a significant level would require further reductions in both the experimental error and theoretical SM uncertainty. Such improvements could lead to stringent tests of \lq\lq slepton universality" of the charged current sector of the MSSM, for which it is often assumed that the left-handed first and second generation sleptons $\widetilde{e}_L$ and $\widetilde{\mu}_L$ are degenerate (see e.g.~\cite{Martin:1997ns}
) and thus $\Delta R_{e/\mu}^{SUSY} \simeq 0$.


In section \ref{sec:rpv}, we consider corrections to $R_{e/\mu}$ from R-parity violating (RPV) processes.  These corrections enter at tree-level, but are suppressed by couplings whose strength is  contrained by other measurements. In order to analyze these constraints, we perform a fit to the current low energy precision observables. We find that, at 95\% C.L., the magnitude of RPV contributions to $\Delta R_{e/\mu}^{SUSY}$ could be several times larger than the combined theoretical and anticipated experimental errors for the  future 
$R_{e/\mu}$  experiments. We summarize the main results and provide conclusions in section \ref{sec:conclude}. 

\section{R-parity conserving interactions }
\label{sec:loops}

\subsection{Pseudoscalar contributions}

The tree-level matrix element for $\pi^+ \to \ell^+ \: \nu_\ell$ that arises from the  $(V-A)\otimes (V-A)$ four fermion operator is 
\begin{eqnarray}
\nonumber
i\mathcal{M}^{(0)}_{AV}& =& -i 2\sqrt{2}\, G_\mu V_{ud} \, \bra{0} {\bar d}\gamma^\lambda P_L\, u\ket{\pi^+}\, \overline{u}_\nu \gamma_\lambda P_L\,  v_{\ell}\\
\label{eq:tree}
 &=&2V_{ud} F_\pi G_\mu m_\ell \: \overline{u}_\nu P_R\,  v_{\ell} \;,
\end{eqnarray}
where $P_{L,R}$ are the left- and right-handed projection operators,  
\begin{equation}
\label{eq:fpiexp}
F_\pi = 92.4\pm 0.07 \pm 0.25\quad {\rm MeV}
\end{equation}
is the pion decay constant, $G_\mu$ is the Fermi constant extracted from the muon lifetime, and $V_{ud}$ is the $(1,1)$ component of the CKM matrix.  The first error in Eq.~(\ref{eq:fpiexp}) is experimental while the second arises from uncertainties associated with QCD effects in the one-loop SM electroweak radiative corrections to the $\pi_{\mu 2}$ decay rate. The superscript \lq\lq ${(0)}$" and subscript \lq\lq $AV$" in Eq.~(\ref{eq:tree}) denote a tree-level,  axial vector contribution. At one-loop order, one must subtract the radiative corrections to the muon-decay amplitude --- since $G_\mu$ is obtained from the muon lifetime --- while adding the corrections to the semileptonic CC amplitude. The corrections to the muon-decay amplitude as well as lepton flavor-independent contributions to the semileptonic radiative corrections cancel from $R_{e/\mu}$.

Now consider the contribution from an induced pseudoscalar four fermion effective operator of the form 
\be
\Delta \mathcal{L}_{PS} = - \frac{G_{PS} V_{ud}}{\sqrt{2}} \: \overline{\nu} (1+\gamma^5) \ell \: \overline{d} \gamma^5 u \;.
\ee
Contributions to $R_{e/\mu}$ from operators of this form were considered in a model-independent operator framework in Ref.~\cite{Campbell:2003ir}.  In the MSSM, such an operator can arise at tree-level (Fig.~\ref{fig:feynman}a) through charged Higgs exchange and at one-loop through box graphs (Fig.~\ref{fig:feynman}d).  These amplitudes determine the value of $G_{PS}$.  The total matrix element is
\be
\label{eq:pseudo}
i \mathcal{M}^{(0)}_{AV} + i \mathcal{M}_{PS} = V_{ud} F_\pi G_\mu m_\ell \: \overline{u}_\nu (1+\gamma^5) v_{\ell} 
 \left[ 1 + \frac{G_{PS}}{G_\mu} \: \omega_\ell \right]
\ee
where 
\be
\omega_\ell \equiv \frac{m_\pi^2}{m_\ell(m_u + m_d)} \simeq \left\{ \ba{ccc} 5\times 10^3 & \; & \ell=e \\
           20 & \; & \ell=\mu \ea \right. \;
\ee
is an enhancement factor reflecting the absence of helicity suppression in 
pseudoscalar contributions as compared to $(V-A)\otimes(V-A)$ contributions \cite{Herczeg:1995kd}.  Pseudoscalar contributions will be relevant to the interpretation of $R_{e/\mu}$  if
\be
\left| \frac{G_{PS}}{G_\mu} \right| \; \omega_\ell \gtrsim 0.0005 \;, \label{eq:pscond} \;
\ee
and if $G_{PS}\, \omega_\ell$ is lepton-flavor dependent. 

The tree-level pseudoscalar contribution (Fig.~\ref{fig:feynman}a) gives
\be
G_{PS}^{(0)} = \frac{m_\ell \tan\beta (m_u \cot\beta - m_d \tan\beta)}{\sqrt{2} m_{H^+}^2 v^2} \;,
\ee
where $m_{H^+}$ is the mass of the charged Higgs boson.  Thus, we have
\be
\frac{G_{PS}^{(0)}}{G_\mu} \: \omega_\ell = \frac{m_\pi^2 \tan\beta (m_u \cot\beta - m_d \tan\beta)}{(m_u+m_d)m_{H^+}^2} \;.
\ee
It is indeed possible to satisfy (\ref{eq:pscond}) for 
\be
\label{eq:hplus}
\tan\beta \; \gtrsim \; 20 \: \left(\frac{m_{H^+}}{100 \; \textrm{GeV}} \right) \;.
\ee
Note that the combination $G_{PS}^{(0)}/G_\mu \times \omega_\ell$ entering Eq.~(\ref{eq:pseudo}) is independent of lepton flavor and will cancel from $R_{e/\mu}$. In principle, however, the extraction of $F_\pi$ from $\pi_{\mu 2}$ decay could be affected by tree-level charged Higgs exchange if the correction in Eq.~(\ref{eq:pscond}) is $\gtrsim 0.003$ in magnitude, corresponding to a shift comparable to the theoretical SM uncertainty as estimated in Ref.~\cite{Marciano:1993sh}. In the case of charged Higgs exchange, one would require $\tan\beta \gtrsim 120\, (m_{H^+}/100 \; \textrm{GeV})$ to generate such an effect. 

\begin{figure}[t]
\begin{center}
\mbox{\hspace*{-0.7cm}\epsfig{file=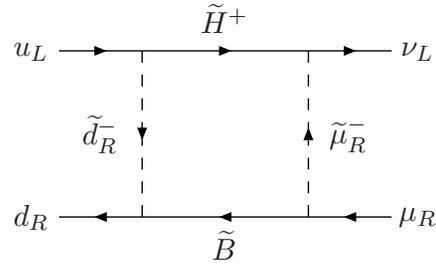,height=3.5cm}}
\end{center}
\caption{\it\small 
This contribution to $G_{PS}^{(1)}$ is suppressed by $\epsilon^2 = y_\mu y_d$.}
\label{fig:minibox}
\end{figure}

One-loop contributions to $G_{PS}$ are generated by box graphs (Fig.~\ref{fig:feynman}d). The magnitude of these contributions is governed by the strength of chiral symmetry breaking in both the quark and lepton sectors. Letting $\epsilon$ generically denote either a Yukawa coupling $y_f$ or a ratio $m_f/M_{SUSY}$ (where $f = e, \; \mu, \; u,$ or $d$), we find that
\be
\frac{G_{PS}^{(1)}}{G_\mu} \sim \frac{\alpha}{8\pi s_W^2} \: \left(\frac{m_W}{M_{SUSY}}\right)^2 \: \epsilon^2 \;, \label{eq:psbox}
\ee
where the superscript ``$(1)$'' denotes one loop induced pseudoscalar interaction.  We have verified by explicit computation that the $\mathcal{O}(\epsilon)$ contributions vanish.  The reason is that in each pair of incoming quarks or outgoing leptons the two fermions must have opposite chirality in order to contribute  to $G_{PS}^{(1)}$.  Since CC interactions in the MSSM are purely left-handed, the chirality must change at least twice in each graph, with each flip generating a factor of $\epsilon$.  For example, we show one pseudoscalar contribution in Fig.~\ref{fig:minibox} that is proportional to $\epsilon^2 = y_\mu y_d$.  Here, the chirality changes at the $u\widetilde{d}\widetilde{H}$ and $\nu\widetilde{\mu}\widetilde{H}$ vertices.  Potentially, this particular contribution can be enhanced for large $\tan\beta$; however, to satisfy (\ref{eq:pscond}), we need
\be
\tan \beta \; \gtrsim \; 10^3 \: \left( \frac{M_{SUSY}}{100 \; \textrm{GeV} } \right)^3 \;.
\ee
These extreme values of $\tan\beta$ can be problematic, leading $y_b$ and $y_\tau$ to become nonperturbatively large.  To avoid this scenario, we need roughly $\tan\beta \lesssim 65$ (see~\cite{Martin:2001st} and references therein).

Pseudoscalar contributions can also arise through mixing of left- and right-handed scalar superpartners.  Since each left-right mixing insertion introduces a factor of $\epsilon$, the leading contributions to $G_{PS}^{(1)}$ will still be $\mathcal{O}(\epsilon^2)$.  However, if the triscalar SUSY-breaking parameters $a_f$ are  not suppressed by $y_f$ as normally assumed, it is possible to have $\epsilon \sim \mathcal{O}(1)$, potentially leading to significant contributions.  This possibility, although not experimentally excluded, is considered theoretically ``unnatural'' as it requires some fine-tuning to avoid spontaneous color and charge breaking (see Ref.~\cite{Profumo:2006yu} for discussion).  Neglecting this possibility and extremely large values of $\tan\beta$, we conclude that loop-induced pseudoscalar contributions are much too small to be detected at upcoming experiments.

\subsection{Axial vector contributions}

One-loop radiative corrections also contribute to the axial vector matrix element.  The total matrix element can be written as
\be
i \mathcal{M}_{AV} = V_{ud} f_\pi G_\mu m_\ell \: \overline{u}_\nu (1+\gamma^5) v_{\ell} 
 \left[ 1 + \Delta {\hat r}_\pi-\Delta {\hat r}_\mu \right] \;,
\ee
where $\Delta {\hat r}_\pi$ and $\Delta {\hat r}_\mu$ denote one-loop contributions to the semileptonic and $\mu$-decay amptlidues, respectively and where the hat indicates quantities renormalized in the modified dimensional reduction ($\overline{DR}$) scheme.  Since $\Delta {\hat r}_\mu$ cancels from $R_{e/\mu}$, we concentrate on the SUSY contributions to $\Delta {\hat r}_\pi$ that do not cancel from $R_{e/\mu}$. It is helpful to distinguish various classes of contributions 
\be
\Delta {\hat r}_\pi^{SUSY} =  \Delta_L^{\ell} + \Delta_V^{\ell}+ \Delta_L^{q} + \Delta_V^{q}+ \Delta_B +\Delta_{GB}\;,
\ee
where  $\Delta_L^{\ell}$ ($\Delta_L^{q}$), $\Delta_V^{\ell}$ ($\Delta_V^{q}$), $\Delta_B$, and $\Delta_{GB}$ denote leptonic (hadronic) external leg (Fig.~\ref{fig:feynman}b), leptonic (hadronic) vertex (Fig.~\ref{fig:feynman}c), box graph (Fig.~\ref{fig:feynman}d), and gauge boson propagator contributions, respectively.  The corrections $\Delta_{L,V}^{q}$ and $\Delta_{GB}$ cancel from $R_{e/\mu}$, so we do not discuss them further (we henceforth omit the \lq\lq $\ell$" superscript). The explicit general formulae for $\Delta_{L,\, V,\, B}$, calculated in $\overline{DR}$, are given in appendix~\ref{appA}.  We have verified that $\Delta_L$ and $\Delta_V$ agree with Ref.~\cite{Katz:1998br} for case of a pure SU(2)$_L$ chargino/neutralino sector.

At face value, it appears from equations (\ref{eq:deltaL}-\ref{eq:deltaB}) that $\Delta R_{e/\mu}^{\textrm{SUSY}}$ carries a non-trivial dependence on MSSM parameters since the SUSY masses enter both explicitly in the loop functions and implicitly in the mixing matrices $Z$, defined in equations (\ref{eq:Zn}-\ref{eq:ZL}).  Nevertheless, we are able to identify a relatively simple dependence on the SUSY spectrum.

We first consider $\Delta R_{e/\mu}^{SUSY}$ in a limiting case obtained with three simplifying assumptions: (1) no flavor mixing among scalar superpartners; (2) no mixing between left- and right-handed scalar superpartners; and (3) degeneracy between $\widetilde{\ell}_L$ and $\widetilde{\nu}_\ell$ and no gaugino-Higgsino mixing.  Our first assumption is well justified; flavor mixing in the slepton and squark sectors is heavily constrained by limits on flavor violating processes, such as $\mu \to e \: \gamma$~\cite{lfv}.  

Our second assumption has minimal impact.  In the absence of flavor mixing, the charged slepton mass matrix decomposes into three $2\times2$ blocks; thus, for flavor $\ell$, the mass matrix in the $\{\widetilde{\ell}_L, \widetilde{\ell}_R\}$ basis is
\be
\left( \begin{array}{cc} 
M_L^2 + \left(s_W^2-\frac{1}{2}\right) m_Z^2 \cos 2\beta & m_\ell \left( \frac{a_\ell}{y_\ell} - \mu \tan \beta \right) \\
m_\ell \left( \frac{a_\ell}{y_\ell} - \mu \tan \beta \right) & M_R^2 - s_W^2 m_Z^2 \cos 2\beta \end{array} \right) \notag \;,
\ee
where $M_L^2$ ($M_R^2$) is the SUSY-breaking mass parameter for left-handed (right-handed) sleptons, $a_\ell$ is the coefficient for the SUSY-breaking triscalar interaction, $y_\ell$ is the Yukawa coupling, and $\mu$ is the Higgsino mass parameter.
Under particular models of SUSY-breaking mediation, it is usually assumed that $a_\ell/y_\ell \sim M_{SUSY}$, and thus left-right mixing is negligible for the first two generations due to the smallness of $m_e$ and $m_\mu$.  Of course, $a_\ell$ could be significantly larger and induce significant left-right mixing ~\cite{Profumo:2006yu}.  For reasons discussed above, we neglect this possibility.

We have adopted the third assumption for purely illustrative purposes; we will relax it shortly.  Clearly, fermions of the same weak isospin doublet are not degenerate; their masses obey
\bea
m_{\widetilde{\ell}_L}^2 &=& m_{\widetilde{\nu}_\ell}^2 - m_W^2 \cos 2\beta + m_\ell^2 
\label{eq:sleptnondeg} \\
m_{\widetilde{d}_L}^2 &=& m_{\widetilde{u}_L}^2 - m_W^2 \cos 2\beta + m_d^2 - m_u^2 \;. 
\label{eq:squarknondeg}
\eea
In addition, gaugino mixing is certainly always present, as the gaugino mass matrices contain off-diagonal elements proportional to $m_Z$ [see Eqs.~(\ref{eq:Zn2}, \ref{eq:Zpm2})].  However, the third assumption becomes valid for $M_{SUSY} \gg m_Z$.

Under our three assumptions, the SUSY vertex and external leg corrections sum to a constant that is independent of the superpartner masses, leading to considerable simplifications.  The Bino [U(1)$_Y$ gaugino] vertex and external leg corrections exactly cancel.  The Wino [SU(2)$_L$ gaugino] vertex and leg corrections do not cancel; rather, $\Delta_V + \Delta_L =  {\alpha}/{4 \pi s_W^2}$, a constant that carries no dependence on the slepton, gaugino, or Higgsino mass parameters.  The occurrence of this constant is merely an artifact of our use of the 
$\overline{\textrm{DR}}$ renormalization scheme.  (In comparison, in modified minimal subtraction, we find $\Delta_V + \Delta_L = 0$ in this same limit.\symbolfootnote[2]{Technically, since $\overline{MS}$ breaks SUSY, it is not the preferred renormalization scheme for the MSSM.  However, this aspect is not important in the present calculation.})  This dependence on renormalization scheme cancels in $R_{e/\mu}$.  (In addition, this scheme-dependent constant enters into the extraction of $G_\mu$; hence, the individual decay widths $\Gamma(\pi \to \ell \nu_\ell)$ are also independent of renormalization scheme.)

The reason for this simplification is that under our assumptions, we have effectively taken a limit that is equivalent to computing the one-loop corrections in the absence of electroweak symmetry breaking.   In the limit of unbroken SU(2)$_L\times$U(1)$_Y$, the one-loop SUSY vertex and external leg corrections sum to a universal constant which is renormalization scheme-dependent, but renormalization scale-independent~\cite{Katz:1998br}.  (For unbroken SU(2)$_L$, the SM vertex and external leg corrections yield an additional logarithmic scale dependence; hence, the SU(2)$_L$ $\beta$-function receives contributions from both charge and wavefunction renormalization.) In addition, virtual Higgsino contributions are negligible, since their interactions are suppressed by small first and second generation Yukawa couplings. Setting all external momenta to zero and working in the limit of unbroken SU(2)$_L$ symmetry, we find that the Higgsino contributions to $\Delta_L+\Delta_V$ are $y_{\ell}^2/32\pi^2$.

In this illustrative limit, the only non-zero contributions to $\Delta R_{e/\mu}^{\textrm{SUSY}}$ come from two classes of box graphs (Fig.~\ref{fig:feynman}d) --- one involving purely Wino-like interactions and the other with both a virtual Wino and Bino.  The sum of these graphs is
\be
\Delta_B^{(\ell)} = \frac{\alpha}{12\pi s_W^2} \left(\frac{m_W^2}{M^2_2} \right)
\left[  F_1(x_L,x_Q) + t_W^2 F_2 (x_B,x_L,x_Q) \right] \label{eq:box2}
\ee
where we have defined
\begin{align}
F_1(x_L,x_Q) & \equiv \; \frac{3}{2} \left[ \frac{ x_L(x_L-2) \ln x_L }{(x_L-x_Q)(1-x_L)^2}  \right. \\ \notag
        & \left. + \frac{ x_Q(x_Q-2) \ln x_Q}{(x_Q-x_L)(1-x_Q)^2} - \frac{1}{(1-x_L)(1-x_Q)} \right] \notag
\end{align}
and
\begin{align}
F_2(x_B,x_L,x_Q) \equiv & \; \frac{1}{2} \: \left[ \frac{x_B(x_B+2\sqrt{x_B}) \ln x_B }{(1-x_B)(x_B-x_L)(x_B-x_Q) } \right. \notag \\ 
                 & \; + \frac{x_L(x_L+2\sqrt{x_B}) \ln x_L }{(1-x_L)(x_L-x_B)(x_L-x_Q) } \\ \notag
                 & \; + \left. \frac{x_Q(x_Q+2\sqrt{x_B}) \ln x_Q }{(1-x_Q)(x_Q-x_L)(x_Q-x_B) }  \right] \;, \notag
\end{align}
where $x_B \equiv M_1^2 / M_2^2$, $x_L \equiv m_{\widetilde{\ell}}^2/M^2_2$, and $x_Q \equiv m_{\widetilde{Q}}^2/M^2_2$, with
masses $M_1$, $M_2$, $m_{\widetilde{\ell}}$, and $m_{\widetilde{Q}}$ of the Bino, Wino, left-handed $\ell$-flavored slepton, and left-handed 1st generation squark, respectively.
Numerically, we find that always $F_1 \gg F_2$; the reason is that the sum of Bino-Wino graphs tend to cancel, while the sum of pure Wino graphs all add coherently.  Hence, Bino exchange (through which the term proportional to $F_2$ arises) does not significantly contribute to $\Delta R_{e/\mu}^{\textrm{SUSY}}$.  

\begin{figure}[!t]
\begin{center}
\mbox{\hspace*{-0.7cm}\epsfig{file=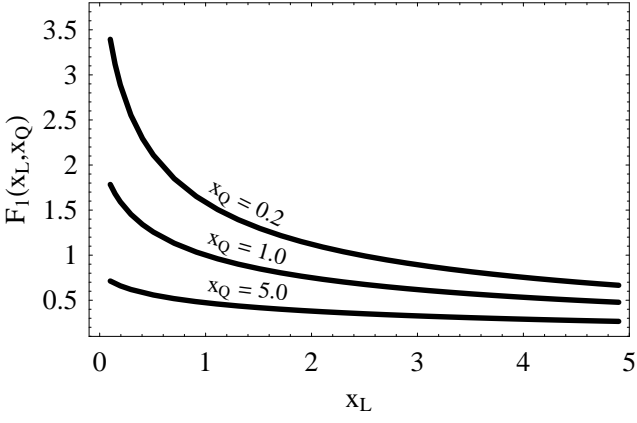,height=5.5cm}}
\end{center}
\caption{\it\small 
The box graph loop function $F_1(x_L,x_Q)$ as a function of $x_L \equiv m_{\widetilde{L}}^2/M_2^2$ for several values of $x_Q \equiv m_{\widetilde{Q}}^2/M^2_2$.  For $x_L \sim x_Q \sim 1$ (i.e. SUSY masses degenerate), $F_1(x_L,x_Q) \sim 1$.  For $x_L \gg 1$ or $x_Q \gg 1$ (i.e. very massive sleptons or squarks), $F_1(x_L,x_Q) \to 0$.}
\label{fig:DPlot}
\end{figure}

In Fig.~\ref{fig:DPlot}, we show $F_1(x_L,x_Q)$ as a function of $x_L$ for fixed $x_Q$.  Since $F_1$ is symmetric under $x_L \leftrightarrow x_Q$, Fig.~\ref{fig:DPlot} also shows $F_1$ as a function of $x_Q$, and hence how $\Delta_B$ depends on $m_{\widetilde{u}_L}$.  For $x_L, \: x_Q \sim 1$, we have $F_1 \sim \mathcal{O}(1)$, while if either $x_L \gg 1$ or $x_Q \gg 1$, then $F_1 \to 0$, which corresponds to the decoupling of heavy sleptons or squarks.  There is no enhancement of $\Delta_B$ for $x_L \ll 1$ or $x_Q \ll 1$ (i.e. if $M_2$ is very heavy) due to the overall $1/M_2^2$ suppression in (\ref{eq:box2}).  

The total box graph contribution is
\bea
\frac{\Delta R_{e/\mu}^{\textrm{SUSY}}}{R_{e/\mu}^{\textrm{SM}}} &=& 2 \; \textrm{Re}[\Delta^{(e)}_B - \Delta^{(\mu)}_B] \notag \\
  &\simeq& \; \frac{\alpha}{6\pi s_W^2} \; \left(\frac{m_W}{M_2} \right)^2 \; \label{eq:box} \\
  &\;& \; \; \times \left[
F_1 \left( \frac{m_{\widetilde{e}}^2}{ M_2^2}, \; \frac{m_{\widetilde{Q}}^2}{ M_2^2} \right) -
F_1 \left( \frac{m_{\widetilde{\mu}}^2}{ M_2^2}, \; \frac{m_{\widetilde{Q}}^2}{ M_2^2} \right) \right] \;. \notag
\eea
Clearly $\Delta R_{e/\mu}^{\textrm{SUSY}}$ vanishes if both sleptons are degenerate and is largest when they are far from degeneracy, such that $m_{\widetilde{e}_L} \gg m_{\widetilde{\mu}_L}$ or $m_{\widetilde{e}_L} \ll m_{\widetilde{\mu}_L}$.  In the latter case, we have 
\be
\left|\frac{\Delta R_{e/\mu}^{\textrm{SUSY}}}{R_{e/\mu}^{\textrm{SM}}}\right| \; \lesssim \; 0.001 \times \left(\frac{100 \; \textrm{GeV}}{M_{SUSY}}\right)^{2} \;
\ee
for e.g. $M_{SUSY} \equiv M_2 \sim m_{\widetilde{u}_L} \sim m_{\widetilde{e}_L} \ll m_{\widetilde{\mu}_L}$.

We now relax our third assumption to allow for gaugino-Higgsino mixing and non-degeneracy of $\widetilde{\ell}$ and $\widetilde{\nu}_\ell$.  Both of these effects tend to spoil the universality of $\Delta_V+\Delta_L$, giving
\be
\Delta_V + \Delta_L-\frac{\alpha}{4\pi s_W^2} = \frac{\alpha}{8\pi s_W^2} \: f \simeq 0.001 \: f \;.
\ee
The factor $f$ measures the departure of $\Delta_V+\Delta_L$ from universality.  If the SUSY spectrum is such that our third assumption is valid, we expect $f \to 0$ . For realistic values of the SUSY parameters, two effects lead to a non-vanishing $f$: (a) splitting between the masses of the charged and neutral left-handed sleptons that results from breaking of SU(2)$_L$, and (b) gaugino-Higgsino mixing. The former effect is typically negligible. To see why, we recall from Eq.~(\ref{eq:sleptnondeg}) that
\be
m_{\widetilde{\ell}} = m_{\widetilde{\nu}_\ell} \; \left[ 1 + \mathcal{O}\left(\frac{m_W^2}{m_{\widetilde{\ell}}^2}\right) \right] \;,
\ee
where we have neglected the small non-degeneracy  proportional to the square of the lepton Yukawa coupling. We find that the leading contribution to $f$ from this non-degeneracy is at least $\mathcal{O}(m_W^4/m_{\widetilde{\ell}}^4)$, which is $\lesssim 0.1$ for $ m_{\widetilde{\ell}}\gtrsim 2M_W$.

\begin{figure}[!t]
\begin{center}
\mbox{\hspace*{-0.7cm}\epsfig{file=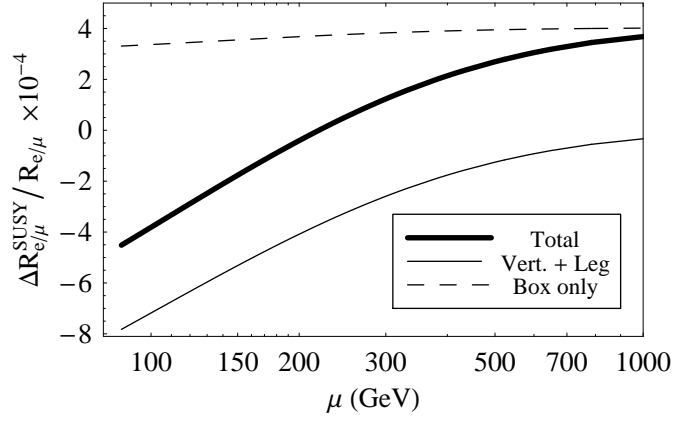,height=5.5cm}}
\end{center}
\caption{\it\small 
$\Delta R_{e/\mu}^{\textrm{SUSY}}$ versus $\mu$, with fixed parameters $M_1= 100$ GeV, $M_2 = 150$ GeV, $m_{\widetilde{e}_L}=100$ GeV, $m_{\widetilde{\mu}_L} = 500$ GeV, $m_{\widetilde{u}_L}=200$ GeV.  Thin solid line denotes contributions from $(\Delta_V+\Delta_L)$ only; dashed line denotes contributions from $\Delta_B$ only; thick solid line shows the sum of both contributions to $\Delta R_{e/\mu}^{\textrm{SUSY}}$.}
\label{fig:VLB1}
\end{figure}

Significant gaugino mixing can induce $f \sim \mathcal{O}(1)$.  The crucial point is that the size of $f$ from gaugino mixing is governed by the size of $M_2$.  If $M_2 \gg m_Z$, then the Wino decouples from the Bino and Higgsino, and contributions to $\Delta_V + \Delta_L$ approach the case of unbroken SU(2)$_L$.  On the other hand, if $M_2 \sim m_Z$, then $\Delta_V + \Delta_L$ can differ substantially from $\alpha/4\pi s_W^2$.

\begin{figure}[!t]
\begin{center}
\mbox{\hspace*{-0.7cm}\epsfig{file=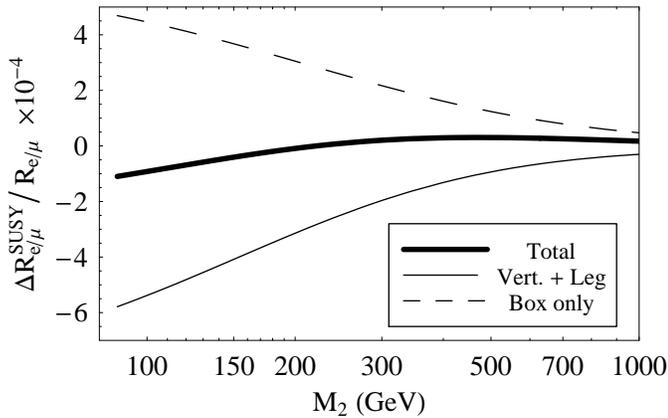,height=5.5cm}}
\end{center}
\caption{\it\small 
$\Delta R_{e/\mu}^{\textrm{SUSY}}/R_{e/\mu}^{SM}$ as a function of $M_2$, with $\mu = 200$ GeV and all other parameters fixed as in Fig.~\ref{fig:VLB1}.  Each line shows the contribution indicated as in the caption of Fig.~\ref{fig:VLB1}. 
}
\label{fig:VLB2}
\end{figure}

In the limit that $m_{\widetilde{\ell}_L} \gg M_2$ ($\ell = e$, $\mu$), we also have a decoupling scenario where $\Delta_B = 0$, $\Delta_V + \Delta_L = \frac{\alpha}{4\pi s_W^2}$, and thus $f=0$.  Hence, a significant contribution to $\Delta R_{e/\mu}$ requires at least one light slepton.  However, regardless of the magnitude of $f$, if $m_{\widetilde{e}_L} = m_{\widetilde{\mu}_L}$, then these corrections will cancel from $R_{e/\mu}$.

It is instructive to consider the dependence of individual contributions $\Delta_B$ and $\Delta_V+\Delta_L$ to $\Delta R_{e/\mu}^{SUSY}$, as shown in Figs.~\ref{fig:VLB1} and \ref{fig:VLB2}.  In Fig. \ref{fig:VLB1}, we plot the various contributions as a function of $\mu$, with $M_1= 100$ GeV, $M_2 = 150$ GeV, $m_{\widetilde{e}_L}=100$ GeV, $m_{\widetilde{\mu}_L} = 500$ GeV, $m_{\widetilde{u}_L}=200$ GeV.  We see that the $\Delta_V+\Delta_L$ contributions (thin solid line) vanish for large $\mu$, since in this regime gaugino-Higgsino mixing is suppressed and there is no $\Delta_V+\Delta_L$ contribution to $\Delta R_{e/\mu}^{SUSY}$.  However, the $\Delta_B$ contribution (dashed line) is nearly $\mu$-independent, since box graphs with Higgsino exchange (which depend on $\mu$) are suppressed in comparison to those with only gaugino exchange.  In Fig. \ref{fig:VLB2}, we plot these contributions as a function of $M_2$, with $\mu= 200$ GeV and all other parameters fixed as above.  We see that both $\Delta_V+\Delta_L$ and $\Delta_B$ contributions vanish for large $M_2$.  

One general feature observed from these plots is that $\Delta_V+\Delta_L$ and $\Delta_B$ contributions tend to cancel one another; therefore, the largest total contribution to $\Delta R_{e/\mu}^{SUSY}$ occurs when either $\Delta_V+\Delta_L$ or $\Delta_B$ is suppressed in comparison to the other.  This can occur in the following ways: (1) if $\mu \gg m_Z$, then $\Delta_B$ may be large, while $\Delta_V+\Delta_L$ is suppressed, and (2) if $m_{\widetilde{u}_L}, \: m_{\widetilde{d}_L} \gg m_Z$, then $\Delta_V+\Delta_L$ may be large, while $\Delta_B$ is suppressed.  In Fig.~\ref{fig:VLB2}, we have chosen parameters for which there is a large cancellation between $\Delta_V+\Delta_L$ and $\Delta_B$.  However, by taking the limits $\mu \to \infty$ or $m_{\widetilde{u}_L}, \; m_{\widetilde{d}_L} \to \infty$, $\Delta R_{e/\mu}^{SUSY}$ would coincide the $\Delta_B$ or $\Delta_V+\Delta_L$ contributions, respectively.  

Because the $\Delta_V+\Delta_L$ and $\Delta_B$ contributions tend to cancel, it is impossible to determine whether $\widetilde{e}_L$ or $\widetilde{\mu}_L$ is heavier from $R_{e/\mu}$ measurements alone.  For example, a positive deviation in $R_{e/\mu}$ can result from two scenarios: (1) $\Delta R_{e/\mu}^{SUSY}$ is dominated by box graph contributions with $m_{\widetilde{e}_L} < m_{\widetilde{\mu}_L}$, or (2) $\Delta R_{e/\mu}^{SUSY}$ is dominated by $\Delta_V+\Delta_L$ contributions with $m_{\widetilde{e}_L} > m_{\widetilde{\mu}_L}$.

Guided by the preceding analysis, we expect for $\Delta R_{e/\mu}^{SUSY}$:
\begin{itemize}
\item The maximum contribution is $\left| \Delta R_{e/\mu}^{SUSY} / R_{e/\mu} \right| \sim 0.001$.
\item Both the vertex + leg and box contributions are largest if $M_2 \sim \mathcal{O}(m_Z)$ and vanish if $M_2 \gg m_Z$.  If $M_2 \sim \mathcal{O}(m_Z)$, then at least one chargino must be light.
\item The contributions to $\Delta R_{e/\mu}^{SUSY}$ vanish if $m_{\widetilde{e}_L}=m_{\widetilde{\mu}_L}$ and are largest if either $m_{\widetilde{\mu}_L} \ll m_{\widetilde{e}_L}$ or $m_{\widetilde{\mu}_L} \gg m_{\widetilde{e}_L}$.
\item The contributions to $\Delta R_{e/\mu}^{SUSY}$ are largest if $\widetilde{e}_L$ or $\widetilde{\mu}_L$ is $\mathcal{O}(m_Z)$.
\item If $\mu \gg m_Z$, then the lack of gaugino-Higgsino mixing suppresses the $\Delta_V+\Delta_L$ contributions to $\Delta R_{e/\mu}^{SUSY}$.  
\item If $m_{\widetilde{u}_L}, \: m_{\widetilde{d}_L} \gg m_Z$, then the $\Delta_B$ contributions to $\Delta R_{e/\mu}^{SUSY}$ are suppressed due to squark decoupling.
\item If $\widetilde{u}_L$, $\widetilde{d}_L$, and $\mu$ are all $\mathcal{O}(m_Z)$, then there may be cancellations between the $\Delta_V + \Delta_L$ and $\Delta_B$ contributions.  $\Delta R_{e/\mu}^{SUSY}$ is largest if it is dominated by \emph{either} $\Delta_V + \Delta_L$ \emph{or} $\Delta_B$ contributions.
\end{itemize}

\begin{figure}[!t]
\begin{center}
\mbox{\hspace*{-0.7cm}\epsfig{file=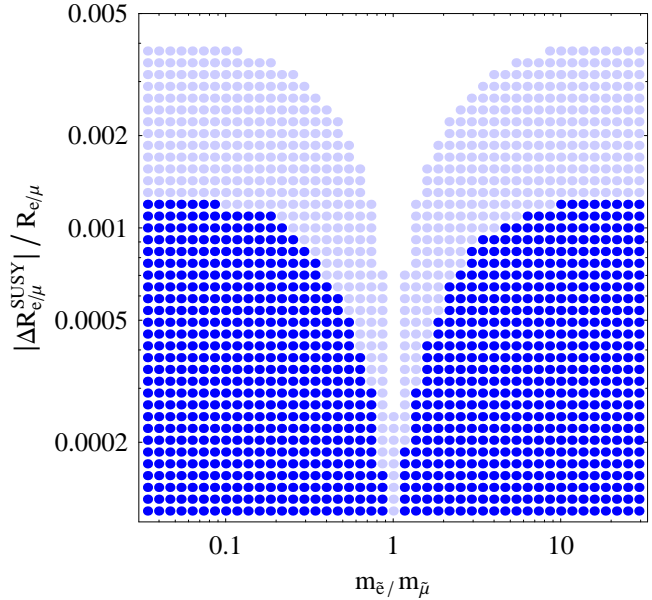,height=8.5cm}}
\end{center}
\caption{\it\small 
$\Delta R_{e/\mu}^{\textrm{SUSY}}$ as a function of the ratio $m_{\widetilde{e}_L}/m_{\widetilde{\mu}_L}$.  Parameter points which obey the LEP II bound are dark blue; parameter points which violate the LEP II bound are light blue.}
\label{fig:massratio}
\end{figure}

We now study $\Delta R_{e/\mu}^{SUSY}$ quantitatively by making a numerical scan over MSSM parameter space, using the following ranges:
\begin{align}
m_Z/2 \;  < \;  \{ M_1, \; |M_2|, \; & |\mu|, \; m_{\widetilde{u}_L} \} \; < \; 1 \; {\textrm{TeV}} \notag \\
m_Z/2 \; < \; \{ m_{\widetilde{\nu}_e}, & \; m_{\widetilde{\nu}_\mu} \} \; < \; 5 \; {\textrm{TeV}} \label{eq:paramrange} \\
1 \; < \; \tan&\:\beta \; < \; 50 \notag \\
{\textrm{sign}}(\mu), \; {\textrm{sign}}&(M_2) = \pm 1 \notag \;, 
\end{align}
where $m_{\widetilde{e}_L}$, $m_{\widetilde{\mu}_L}$, and $m_{\widetilde{d}_L}$ are determined from Eqs. (\ref{eq:sleptnondeg},\ref{eq:squarknondeg}).

\begin{figure}[!t]
\begin{center}
\mbox{\hspace*{-0.7cm}\epsfig{file=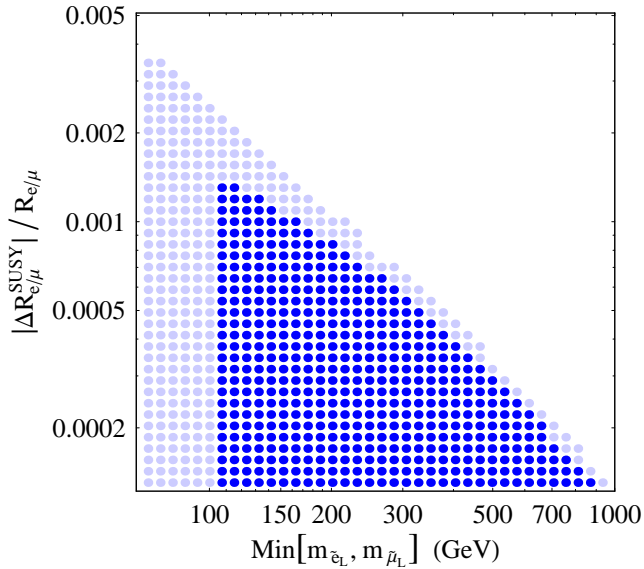,height=8.5cm}}
\end{center}
\caption{\it\small 
$\Delta R_{e/\mu}^{\textrm{SUSY}}$ as a function of Min[$m_{\widetilde{e}_L}$, $m_{\widetilde{\mu}_L}$], the mass of the lightest first or second generation charged slepton. Parameter points which obey the LEP II bound are dark blue; parameter points which violate the LEP II bound are light blue.}
\label{fig:mslep}
\end{figure}

Direct collider searches impose some constraints on the parameter space. Although the detailed nature of these constraints depend on the adoption of various assumptions and on interdependencies on the nature of the MSSM and its spectrum~\cite{Yao:2006px},  we implement them in a  coarse way in order to identify the general trends in corrections to $R_{e/\mu}$.  First, we include only parameter points in which there are no SUSY masses lighter than $m_Z/2$.  (However, the current bound on the mass of lightest neutralino is even weaker than this.)  Second, parameter points which have no charged SUSY particles lighter than 103 GeV are said to satisfy the ``LEP II bound.''  (This bound may also be weaker in particular regions of parameter space.) 

Additional constraints arise from precision electroweak data. We consider only MSSM parameter points whose contributions to oblique parameters S, T, and U agree with Z-pole measurements at 95\% CL~\cite{Yao:2006px}.  Because we have neglected the 3rd generation and right-handed scalar sectors in our analysis and parameter scan, we do not calculate the entire MSSM contributions to S, T, and U.  Rather, we only include those from charginos, neutralinos, and 1st generation left-handed scalar superpartners.  Although incomplete, this serves as a conservative lower bound; in general, the contributions to S, T, and U from the remaining scalar superpartners (that we neglect) only causes further deviations from the measured values of the oblique parameters.  In addition, we include only the lightest CP-even Higgs boson with mass $m_h = 114.4$ GeV, neglecting the typically small contributions from the remaining heavier Higgs bosons.

We do not impose other electroweak constraints in the present study, but note that they will generally lead to further restrictions. For example, the results of the E821 measurement of the muon anomalous magnetic moment~\cite{Bennett:2006fi} tend to favor a positive sign for the $\mu$ parameter and relatively large values of $\tan\beta$. Eliminating the points with $\textrm{sign}(\mu)=-1$ will exclude half the parameter space in our scan, but the general trends are unaffected. 

\begin{figure}[!t]
\begin{center}
\mbox{\hspace*{-0.7cm}\epsfig{file=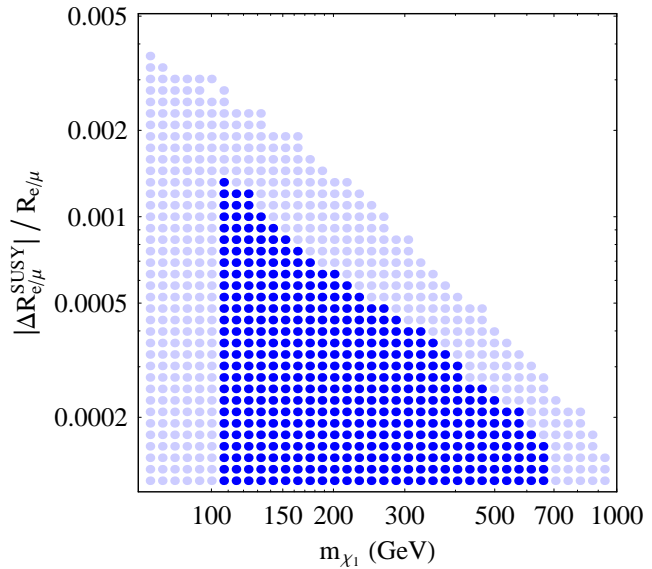,height=8.5cm}}
\end{center}
\caption{\it\small 
$\Delta R_{e/\mu}^{\textrm{SUSY}}$ versus $m_{\chi_1}$, the mass of the lightest chargino. Parameter points which obey the LEP II bound are dark blue; parameter points which violate the LEP II bound are light blue.}
\label{fig:mchi1}
\end{figure}

\begin{figure}[!t]
\begin{center}
\mbox{\hspace*{-0.7cm}\epsfig{file=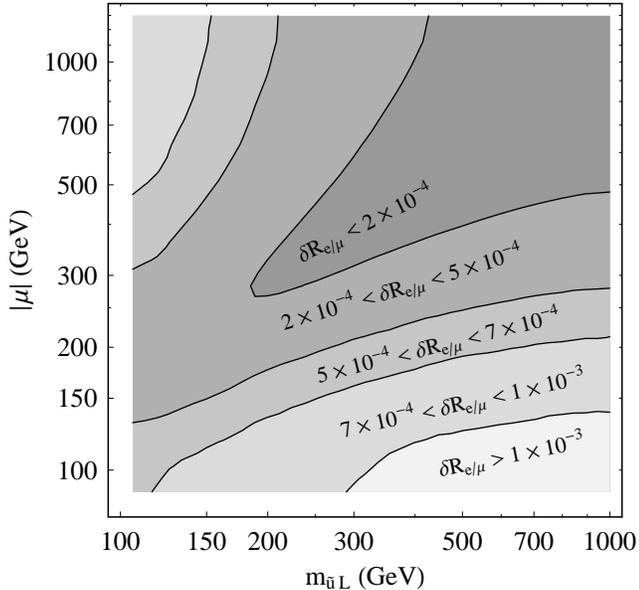,height=8.5cm}}
\end{center}
\caption{\it\small 
Contours indicate the largest values of $\delta R_{e/\mu}^{\textrm{SUSY}}$ obtained by our numerical parameter scan (\ref{eq:paramrange}), as a function of $|\mu|$ and $m_{\widetilde{u}_L}$.  The solid shaded regions correspond to the largest values of $\delta R_{e/\mu}^{\textrm{SUSY}}$ within the ranges indicated.  All values of $\delta R_{e/\mu}^{\textrm{SUSY}}$ correspond to parameter points which satisfy the LEP II bound.
}
\label{fig:contour}
\end{figure}

We show the results of our numerical scan in Figs.~\ref{fig:massratio}--\ref{fig:contour}.  Parameter points which satisfy the LEP II bound are dark blue; those which do not are light blue. In Fig.~\ref{fig:massratio}, we show $\Delta R_{e/\mu}^{SUSY} / R_{e/\mu}$ as a function of the ratio of slepton masses $m_{\widetilde{e}_L}/m_{\widetilde{\mu}_L}$.  If both sleptons are degenerate, then $\Delta R_{e/\mu}^{SUSY}$ vanishes.  Assuming the LEP II bound, in order for a deviation in $R_{e/\mu}$ to match the target precision at upcoming experiments, we must have
\be
\delta R_{e/\mu} \equiv \left|\Delta R_{e/\mu}^{SUSY} / R_{e/\mu}\right| \gtrsim 0.0005 \;, \label{eq:relevant}
\ee
and thus $m_{\widetilde{e}_L}/m_{\widetilde{\mu}_L} \gtrsim 2$ or $m_{\widetilde{\mu}_L}/m_{\widetilde{e}_L} \gtrsim 2$.

In Fig.~\ref{fig:mslep}, we show $\Delta R_{e/\mu}^{SUSY} / R_{e/\mu}$ as a function of Min[$m_{\widetilde{e}_L}$, $m_{\widetilde{\mu}_L}$], the mass lightest first or second generation slepton.  If the lighter slepton is extremely heavy, then both heavy sleptons decouple, causing $\Delta R_{e/\mu}^{SUSY}$ to vanish.  Assuming the LEP II bound, to satisfy (\ref{eq:relevant}), we must have $m_{\widetilde{e}_L} \lesssim 300$ GeV or $m_{\widetilde{\mu}_L} \lesssim 300$ GeV.

In Fig.~\ref{fig:mchi1}, we show $\Delta R_{e/\mu}^{SUSY} / R_{e/\mu}$ as a function of $m_{\chi1}$, the lightest chargino mass.  If $m_{\chi1}$ is large, $\Delta R_{e/\mu}^{SUSY}$ vanishes because $M_2$ must be large as well, suppressing $\Delta_B$ and forcing $\Delta_V$ and $\Delta_L$ to sum to the flavor independent constant discussed above.  Assuming the LEP II bound, to satisfy (\ref{eq:relevant}), we must have $m_{\chi 1} \lesssim 250$ GeV.

Finally, we illustrate the interplay between $\Delta_V+\Delta_L$ and $\Delta_B$ by considering $\delta R^{SUSY}_{e/\mu}$ as a function of $|\mu|$ and $m_{\widetilde{u}_L}$.  In Fig.~\ref{fig:contour}, we show the largest values of $\delta R^{SUSY}_{e/\mu}$ obtained in our numerical parameter scan, restricting to parameter points which satisfy the LEP II bound.  The solid shaded areas correspond to regions of the $|\mu|$-$m_{\widetilde{u}_L}$ plane where the largest value of $\delta R^{SUSY}_{e/\mu}$ lies within the indicated ranges.  It is clear that $\delta R^{SUSY}_{e/\mu}$ can be largest in the regions where either (1) $\mu$ is small, $m_{\widetilde{u}_L}$ is large, and the largest contributions to $\Delta R^{SUSY}_{e/\mu}$ are from $\Delta_V+\Delta_L$, or (2) $\mu$ is large, $m_{\widetilde{u}_L}$ is small, and the largest contribution to $\Delta R^{SUSY}_{e/\mu}$ is from $\Delta_B$.  If both $\mu$ and $m_{\widetilde{u}_L}$ are light, then $\Delta R^{SUSY}_{e/\mu}$ can still be very small due to cancellations, even though both $\Delta_V+\Delta_L$ and $\Delta_B$ contributions are large individually.  More precisely, to satisfy (\ref{eq:relevant}), we need either $\mu \lesssim 250$ GeV, or $\mu \gtrsim 300$ GeV and $m_{\widetilde{u}_L} \lesssim 200$ GeV.

\section{Contributions from R-parity Violating Processes}
\label{sec:rpv}

\begin{figure}[ht]
\begin{center}
\resizebox{3 in}{!}
{\includegraphics*[50,560][290,630]{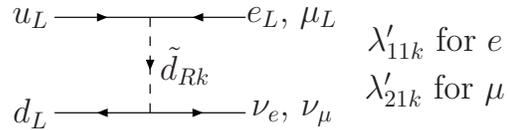}}
\caption{
Tree-level RPV contributions to  $R_{e/\mu}$.
}
\label{fig:RPV}
\end{center}
\end{figure}

In the presence of RPV interactions, tree-level exchanges of sfermions (shown in Fig.~\ref{fig:RPV}), lead to violations of lepton universality and non-vanishing effects in $R_{e/\mu}$.  The magnitude of these tree-level contributions is governed by both the sfermion masses and by the parameters $\lambda_{11k}^{\prime}$ and $\lambda_{21k}^{\prime}$ that are the
coefficients  in RPV interactions:
\begin{equation}
\label{eq:rpv-super}
{\cal L}_{RPV, \ \Delta{L}=1} =\lambda_{ijk}^{\prime} L_i Q_j\tilde{\bar{d}}^\dagger_k
 +\ldots
\end{equation}
Defining~\cite{Ramsey-Musolf:2000qn, Barger:2000gv}
\begin{equation}
\label{eq:deltas}
\Delta_{ijk}^{\prime}(\tilde f)={|\lambda_{ijk}^{\prime}|^2\over 4\sqrt{2}G_\mu
m_{\tilde f}^2}\ge 0,
\end{equation}
contributions to $R_{e/\mu}$ from RPV interactions are
\begin{equation}
\frac{\Delta R_{e/\mu}^{RPV}}{R_{e/\mu}^{SM}}=2 \Delta_{11k}^\prime-2 \Delta_{21k}^\prime.
\end{equation}
Note that RPV contribution to the muon lifetime  (and, thus, the Fermi constant $G_{\mu}$) cancels in $R_{e/\mu}$, 
therefore does not enter the expression.

The quantities $\Delta_{ijk}^\prime$ {\it etc.}  are constrained by existing precision measurements and rare decays.  A summary of the low energy constraints is given in 
Table III of Ref.~\cite{Ramsey-Musolf:2006vr}, which includes tests of CKM unitarity (primarily through RPV effects in superallowed  nuclear
$\beta$-decay that yields a precise value of $|V_{ud}|$ \cite{Hardy:2004id}), atomic
parity violating (PV) measurements of the cesium weak charge $Q_W^{\rm Cs}$ \cite{Ben99},
the ratio $R_{e/\mu}$ itself \cite{Britton:1992xv, Czapek:1993kc}, a
comparison of the Fermi constant $G_\mu$ with the appropriate
combination of $\alpha$, $m_Z$, and $\sin^2\theta_W$ \cite{marciano99}, 
and the electron weak charge determined from SLAC E158 measurement of parity violating M\o ller scattering\cite{E158}.

\begin{figure}[ht!]
\begin{center}
\resizebox{3 in}{!}
{\includegraphics*{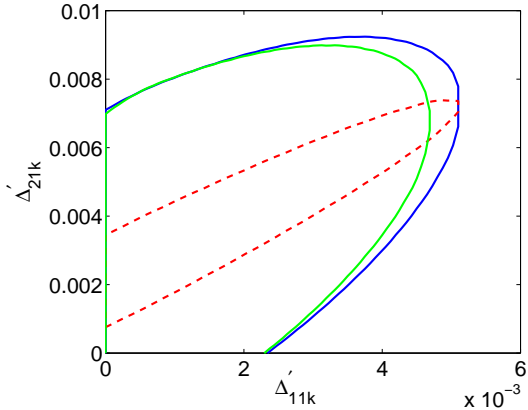}}
\caption{
Present 95\% C.L. constraints on RPV parameters $\Delta^\prime_{j1k}$, $j=1,2$ that enter 
$R_{e/\mu}$ obtained from a fit to precision electroweak observables. Interior of the dark blue contour
 corresponds to the fit using the current value of $\Delta R_{e/\mu}/R_{e/\mu}^{SM}$ \cite{Britton:1992xv, Czapek:1993kc}, while the dashed red contour corresponds to the fit using the future expected experimental precision \cite{triumfproposal}, assuming the same central value.  The light green curve indicates prospective impact of a future measurement of the proton weak charge at Jefferson Lab~\cite{jlab:qweak}.}
\label{fig:rpvcontour}
\end{center}
\end{figure}

In Fig.~\ref{fig:rpvcontour} we show the present 95\% C.L. constraints on the quantities $\Delta_{11k}^\prime$ and $\Delta_{21k}^\prime$ obtained from the aforementioned observables (interior of the blue curve). Since the $\Delta^\prime_{ijk}$ are positive semidefinite quantities, only the region of the contour in the upper right hand quadrant are shown. The green curve indicates the possible implication of a future measurement of the proton weak charge planned at Jefferson Lab~\cite{jlab:qweak}, assuming agreement with the Standard Model prediction for this quantity and the anticipated experimental uncertainty. The dashed red curve shows the possible impact of future measurements of $R_{e/\mu}$, assuming agreement with the present central value but an overall error reduced to the level anticipated in Ref.~\cite{triumfproposal}; with the error anticipated in Ref.~\cite{psiproposal} the width of the band would be a factor of two smaller than shown. 

Two general observations emerge from Fig.~\ref{fig:rpvcontour}. First, given the present constraints, values of $\Delta^\prime_{21k}$ and $\Delta^\prime_{11k}$ differing substantially from zero are allowed. For values of these quantities inside the blue contour,  $\Delta R_{e/\mu}^{SUSY}$ could differ from zero by up to five standard deviations for the error anticipated in Ref.~\cite{triumfproposal}. Such RPV effects could, thus, be considerably larger than the SUSY loop corrections discussed above. On the other hand, agreement of $R_{e/\mu}$ with the SM would lead to considerable tightening of the constraints on this scenario, particularly in the case of $\Delta^\prime_{21k}$, which is currently constrained only by $R_{e/\mu}$ and deep inelastic $\nu$ ($\bar\nu$) scattering~\cite{Zeller:2001hh}. 

The presence of RPV interactions would have significant implications for both neutrino physics and cosmology. It has long been known, for example, that the existence of $\Delta L=\pm 1$ interactions --- such as those that could enter $R_{e/\mu}$ ---  will induce a Majorana neutrino mass~\cite{Schechter:1980gr}, while the presence of non-vanishing RPV couplings would imply that the lightest supersymmetric particle is unstable and, therefore, not a viable candidate for cold dark matter. The future measurements of $R_{e/\mu}$ could lead to substantially tighter constraints on these possibilities or uncover a possible indication of RPV effects. In addition, we note that the present uncertainty associated with RPV effects entering the $\pi_{\mu 2}$ decay rate would affect the value of $F_\pi$ at a level about half the theoretical SM uncertainty as estimated by Ref.~\cite{Marciano:1993sh}.

\section{Conclusions}
\label{sec:conclude}

Given the prospect of two new studies of lepton universality in $\pi_{\ell 2}$ decays~\cite{triumfproposal,psiproposal} with experimental errors that are substantially smaller than for existing measurements and possibly approaching the $5\times 10^{-4}$ level, an analysis of the possible implications for supersymmetry is a timely exercise. In this study, we have considered SUSY effects on the ratio $R_{e/\mu}$ in the MSSM both with and without R-parity violation. Our results indicate that in the R-parity conserving case, effects from SUSY loops can be of order the planned experimental error in particular, limited regions of the MSSM parameter space. Specifically, we find that a deviation in $R_{e/\mu}$ due to the MSSM at the level of
\be
0.0005 \lesssim \left| \frac{\Delta R_{e/\mu}^{SUSY}}{R_{e/\mu}}\right| \lesssim 0.001 \;,
\ee
implies (1) the lightest chargino $\chi_{1}$ is sufficiently light
\be
m_{\chi1} \; \lesssim \;  250 \; \textrm{GeV}\ \ \ , \notag 
\ee
(2) the left-handed selectron $\widetilde{e}_L$ and smuon $\widetilde{\mu}_L$ are highly non-degenerate:
\be
\frac{m_{\widetilde{e}_L}}{m_{\widetilde{\mu}_L}} \; \gtrsim \; 2 \quad \textrm{or} \quad \frac{m_{\widetilde{e}_L}}{m_{\widetilde{\mu}_L}} \; \lesssim \; \frac{1}{2} \notag \;,
\ee
(3) at least one of $\widetilde{e}_L$ or $\widetilde{\mu}_L$ must be light, such that
\bea
m_{\widetilde{e}_L} \lesssim 300 \; \textrm{GeV} \quad \textrm{or} \quad m_{\widetilde{\mu}_L} \lesssim 300 \; \textrm{GeV}, \notag \\
\notag
\eea
and (4) the Higgsino mass parameter $\mu$ and left-handed up squark mass $m_{\widetilde{u}_L}$ satisfy either 
\be
|\mu| \lesssim 250 \; \textrm{GeV} \notag 
\ee
or
\be
|\mu| \gtrsim 300 \; \textrm{GeV}, \; m_{\widetilde{u}_L} \lesssim 200 \; \textrm{GeV}. \notag
\ee
Under these conditions, the magnitude $\Delta R_{e/\mu}^{SUSY}$ may fall within the sensitivity of the new $R_{e/\mu}$ measurements.    

In conventional scenarios for SUSY-breaking mediation, one expects the left-handed slepton masses to be comparable, implying no substantial corrections to SM predictions for $R_{e/\mu}$. Significant reductions in both experimental error and theoretical, hadronic physics uncertainties in $R_{e/\mu}^{SM}$ would be needed to make this ratio an effective probe of the superpartner spectrum.

On the other hand, constraints from existing precision electroweak measurements leave considerable latitude for observable effects from tree-level superpartner exchange in the presence of RPV interactions. The existence of such effects would have important consequences for both neutrino physics and cosmology, as the presence of the $\Delta L\not=0$ RPV interactions would induce a Majorana mass term for the neutrino and allow the lightest superpartner to decay to SM particles too rapidly to make it a viable dark matter candidate. Agreement between the results of the new $R_{e/\mu}$ measurements with $R_{e/\mu}^{SM}$ could yield significant new constraints on these possibilities.

\begin{acknowledgments}

We would like to thank M. Wise for useful discussions.  MRM and ST are supported under U.S Department of Energy Contract \# DE-FG02-05ER41361 and NSF Award PHY-0555674.  SS is supported under U.S Department of Energy Contract \# DE-FG02-04ER-41298.

\end{acknowledgments}

\begin{widetext}

\appendix

\section{\label{appA}General Radiative Corrections in the MSSM}
The MSSM Lagrangian and Feynman rules~\cite{Rosiek:1989rs} are expressed in terms of chargino and neutralino mixing matrices $Z_\pm$ and $Z_N$, respectively, which diagonalize the superpartner mass matrices, defined as follows.  The four neutralino mass eigenstates $\chi^0_i$ are related to the gauge eigenstates $\psi^0 \equiv (\widetilde{B}, \widetilde{W}^3, \widetilde{H}^0_d, \widetilde{H}^0_u)$ by
\be
\psi^0_i = Z_N^{ij} \: \chi^0_j \;, \label{eq:Zn}
\ee
where 
\be
Z_N^T \: \left( \begin{array}{cccc} M_1 & 0 & -c_\beta \: s_W m_Z & s_\beta \:s_W m_Z \\
                                     0  & M_2 & c_\beta \:c_W m_Z & -s_\beta \:c_W m_Z \\
                                    -c_\beta \:s_W m_Z & c_\beta \:c_W m_Z & 0 & -\mu \\
                                    s_\beta \:s_W m_Z & -s_\beta \:c_W m_Z & -\mu & 0 \end{array} \right) Z_N =
\left( \begin{array}{cccc} m_{\chi^0_1} & 0 & 0 & 0 \\
                           0 & m_{\chi^0_2} & 0 & 0 \\
                           0 & 0 & m_{\chi^0_3} & 0 \\
                           0 & 0 & 0 & m_{\chi^0_4} \end{array} \right) \label{eq:Zn2}\;
\ee
is the diagonalized neutralino mass matrix.  The chargino mass eigenstates $\chi^\pm_i$ are related to the gauge eigenstates $\psi^+ \equiv (\widetilde{W}^+, \widetilde{H}^+_u)$ and $\psi^- \equiv (\widetilde{W}^-, \widetilde{H}^-_d)$ by
\be
\psi^\pm_i = Z_\pm^{ij} \: \chi^\pm_j \;,
\ee
where
\be
Z_-^T \: \left( \begin{array}{cc} M_2 & \sqrt{2} s_\beta m_W \\
                                  \sqrt{2} c_\beta m_W & \mu \end{array} \right) \: Z_+ = 
\left( \begin{array}{cc} m_{\chi_1} & 0 \\
                         0 & m_{\chi_2} \end{array} \right) \label{eq:Zpm2}
\ee
is the diagonalized chargino mass matrix.  We note that the off-diagonal elements which mix gauginos and Higgsinos stem solely from electroweak symmetry breaking.

The charged slepton mass eigenstates $\widetilde{L}_i$ are related to the gauge eigenstates $\widetilde{\ell} \equiv (\widetilde{e}_L, \widetilde{\mu}_L,\widetilde{\tau}_L,\widetilde{e}_R,\widetilde{\mu}_R,\widetilde{\tau}_R)$ by
\be
\widetilde{\ell}_i = Z_L^{ij} \: \widetilde{L}_j \;,
\ee
where
\be
Z_L^\dagger \: {\bf M^2_{\widetilde{\ell}} } \: Z_L = 
 \left( \begin{array}{ccc} m^2_{\widetilde{L}_1} & \:       &   0 \\
                                             \:  & \ddots   & \:  \\
                           0                     & \:       & m^2_{\widetilde{L}_6} \end{array} \right)\label{eq:ZL}
\ee
is the diagonalized slepton mass matrix.  There are two classes of off-diagonal elements in ${\bf M^2_{\widetilde{\ell}}}$ which can contribute to slepton mixing: mixing between flavors and mixing between left- and right-handed components of a given flavor, both of which arise through SUSY-breaking terms.  (Left-right mixing due to SUSY-preserving terms will be suppressed by $m_\ell / m_{\widetilde{\ell}}$ and is irrelevant for the first two generations.)  

Similarly, up-type squarks, down-type squarks, and sneutrinos have mixing matrices $Z_U$, $Z_D$, and $Z_\nu$, respectively, defined identically to $Z_L$ --- except for the fact that there are no right-handed sneutrinos in the MSSM and thus there are only three sneutrino mass eigenstates.  

There are three types of contributions to $\Delta R_{e/\mu}^{SUSY}$ in the MSSM: external leg, vertex, and box graph radiative corrections.  The leptonic external leg corrections (Fig.~\ref{fig:feynman}b) are
\begin{align}
\Delta_L^{(i)} = - \frac{\alpha}{16\pi s_W^2} &\left( \frac{}{}\right.  |Z_N^{1j} t_W - Z_N^{2j}|^2 \;
             B(m_{\chi^0_j},m_{\widetilde{\nu}_i}) \label{eq:deltaL}
          + \; 2 \; |Z_-^{1k}|^2 \; B(m_{\chi_k},m_{\widetilde{L}_i}) \\
          &+ \;  |Z_N^{1j} t_W + Z_N^{2j}|^2  \; B(m_{\chi^0_j},m_{\widetilde{L}_i}) 
          + \; 2 \; |Z_+^{1k}|^2 \; B(m_{\chi_k},m_{\widetilde{\nu}_i}) \left. \frac{}{} \right) \;, \notag
\end{align}
where the loop function is~\cite{Passarino:1978jh}

\be
B(m_1,m_2) = \int^1_0 dx \: x \: \ln \left( \frac{ M^2 }{m_1^2 (1-x) + m_2^2 x} \right)\;.\notag
\ee
The leptonic vertex corrections (Fig.~\ref{fig:feynman}c) are
\begin{align}
\Delta_V^{(I)} = \frac{\alpha}{8\pi s_W^2} \left( \frac{}{} \right. & 
             (Z_N^{1j} t_W + Z_N^{2j})\: (Z_N^{1j*} t_W - Z_N^{2j*}) \; C_2(m_{\widetilde{\nu}_i},m_{\chi^0_j},m_{\widetilde{L}_i})\label{eq:deltaV}\\
+ & \; 2  \; (Z_N^{2j*} - t_W \: Z_N^{1j*}) \; Z_+^{1k} \; 
  \left[ \frac{}{} \right. (Z_N^{2j} Z_+^{1k} - \frac{1}{\sqrt{2}} Z_N^{4j} Z_+^{2k})
       \; C_2(m_{\chi^0_j},m_{\widetilde{\nu}_i},m_{\chi_k}) \notag \\
& \qquad \qquad \qquad \qquad \qquad \qquad + \; (Z_N^{2j*} Z_-^{1k} + \frac{1}{\sqrt{2}} Z_N^{3j*} Z_-^{2k}) \; 
   \;  m_{\chi^0_j} m_{\chi_k} \; C_1(m_{\chi^0_j},m_{\widetilde{\nu}_i},m_{\chi_k})\left. \frac{}{}\right] \notag \\
+ & \; 2  \: (Z_N^{2j} + t_W \:  Z_N^{1j}) \; Z_-^{1k} \; 
  \left[ \frac{}{} \right. (Z_N^{2j*} Z_-^{1k} + \frac{1}{\sqrt{2}} Z_N^{3j*} Z_-^{2k})
       \; C_2(m_{\chi_k},m_{\widetilde{L}_i},m_{\chi^0_j})  \notag \\
& \qquad \qquad \qquad \qquad \qquad \qquad + \; (Z_N^{2j} Z_+^{1k} - \frac{1}{\sqrt{2}} Z_N^{4j} Z_+^{2k}) \; 
   \;  m_{\chi^0_j} m_{\chi_k} \; C_1(m_{\chi_k},m_{\widetilde{L}_i},m_{\chi^0_j})\left. \frac{}{}\right] \left. \frac{}{}\right) \notag  \;,
\end{align}
with loop functions
\begin{align}
C_1(m_1,m_2,m_3) & = \int^1_0 dx \: dy \: \frac{1}{m_1^2 x + m_2^2 y + m_3^2(1-x-y) } \notag\\
C_2(m_1,m_2,m_3) & = \int^1_0 dx \: dy \: \ln \left( \frac{M^2}{m_1^2 x + m_2^2 y + m_3^2(1-x-y)} \right) \notag\;.
\end{align}
The corrections from box graphs (Fig.~\ref{fig:feynman}d) are
\begin{align}
\Delta_B^{(I)} = \frac{\alpha m_W^2}{8 \pi s_W^2} \left( \frac{}{} \right. & |Z_-^{1k}|^2
 \; (Z_N^{2m*} + t_W Z_N^{1m*} )\;( Z_N^{2m}-\frac{1}{3} t_W Z_N^{1m} ) \;  \label{eq:deltaB} \; D_1(m_{\chi^0_m},m_{\widetilde{d}_L},m_{\chi_k},m_{\widetilde{L}_i}) \\
+ \; & |Z_+^{1j}|^2
 \; (Z_N^{2m} - t_W Z_N^{1m} )\;( Z_N^{2m*}+\frac{1}{3} t_W Z_N^{1m*} ) \;   
 D_1(m_{\chi_j},m_{\widetilde{u}_L},m_{\chi_m^0},m_{\widetilde{\nu}_i}) \notag  \\
+ \; & Z_-^{1j} Z_+^{1j} 
      \; (Z_N^{2m} - t_W Z_N^{1m} )\;( Z_N^{2m}-\frac{1}{3} t_W Z_N^{1m} )
\; m_{\chi_m^0} m_{\chi_j} \; D_2(m_{\chi_m^0},m_{\widetilde{d}_L},m_{\chi_j},m_{\widetilde{\nu}_i})  \notag \\
+ \; & Z_-^{1k} Z_+^{1k} 
      \; (Z_N^{2m*} + t_W Z_N^{1m*} )\;( Z_N^{2m*}+\frac{1}{3} t_W Z_N^{1m*} )  
\; m_{\chi_m^0} m_{\chi_k} \; D_2(m_{\chi_k},m_{\widetilde{u}_L},m_{\chi_m^0},m_{\widetilde{L}_i}) 
 \left. \frac{}{} \right) \notag  \;,
\end{align}
with loop functions
\be
D_n(m_1,m_2,m_3,m_4) = \int^1_0 dx \: dy \: dz \; \frac{1}{[m_1^2 x + m_2^2 y + m_3^2 z + m_4^2 (1-x-y-z)]^n} \;.\notag
\ee

In formulae (\ref{eq:deltaL}-\ref{eq:deltaB}), $I=1$ corresponds to $\pi \to e \: \nu_e$ and $I=2$ corresponds to $\pi \to \mu \: \nu_\mu$.  All other indeces are summed over.  We use $\overline{\textrm{DR}}$ renormalization at scale $M$.  We have defined $t_W \equiv \tan \theta_W$ and $s_W \equiv \sin \theta_W$.  We have neglected terms proportional to either Yukawa couplings or external momenta (which will be suppressed by ${\mathcal{O}}(m_\pi/M_{SUSY})$). Finally, the SUSY contribution to $R_{e/\mu}$ is
\be
\frac{ \Delta R_{e/\mu}^{\textrm{SUSY}} }{ R_{e/\mu} } 
 = 2 \; \textrm{Re}[\Delta^{(1)}_V - \Delta^{(2)}_V + \Delta^{(1)}_L - \Delta^{(2)}_L + \Delta^{(1)}_B - \Delta^{(2)}_B] \;.
\ee

In addition, the following are some useful formulae needed to show the cancellations of vertex and leg corrections in the limit of no superpartner mixing:
\begin{align}
C_2(&m_1,m_2,m_1) = B(m_2,m_1) \notag \\
2 m_1^2 \: C_1(m_1,m_2,m_1&) - 2 \: B(m_1,m_2) + 2 \: B(m_2,m_1) = 1 \;. \notag  
\end{align}

\end{widetext}

\end{document}